\documentstyle[11pt]{article}
\newcommand{\be}{\begin{equation}}
\newcommand{\ee}{\end{equation}}
\newcommand{\ba}{\begin{eqnarray}}
\newcommand{\ea}{\end{eqnarray}}
\newcommand{\bann}{\begin{eqnarray*}}
\newcommand{\eann}{\end{eqnarray*}}

\begin{document}
\hbadness=10000
\setcounter{page}{1}

\title{
\vspace{-3.0cm}
\hspace{-2.0cm}
\hspace*{\fill} 
\hspace*{\fill}
{\normalsize submitted to Physical Review C} \\*[1.0ex] 
{\huge \bf 
Probing the equation of state in the AGS energy range with 3-d 
hydrodynamics}}
\author{
N. Arbex$^1$\thanks{E. Mail: ARBEX@MAILER.UNI-MARBURG.DE},
U.Ornik$^2$\thanks{E. Mail: ORNIK@WARP.SOULTEK.DE},
M. Pl\"umer$^1$\thanks{E. Mail: PLUEMER@MAILER.UNI-MARBURG.DE} and
R.M. Weiner$^1$\thanks{E. Mail: WEINER@MAILER.UNI-MARBURG.DE}}
\date{$^1$ Physics Department, Univ. of Marburg, Marburg, Germany \\
      $^2$ Soultek Internet Service, Marburg, Germany}
\maketitle
\begin{abstract}
The effect of (i) the phase transition between a quark gluon plasma 
(QGP) and a hadron gas and (ii) the number 
of resonance degrees of freedom
in the hadronic phase
on the single inclusive distributions of 16 different
types of produced hadrons for $Au+Au$ collisions
at $AGS$ energies is studied. 
We have used an exact numerical solution of the relativistic
hydrodynamical equations {\it without free parameters}
which, because of its 3-d character, constitutes a
considerable improvement over the classical Landau solution.
Using two different equations of state (eos) - one containing
a phase transition from QGP to the Hadronic Phase and 
two versions of a purely hadronic eos - we find 
that the first one gives an overall
better description of
the $Au+Au$ experimental data at $AGS$ energies.
We reproduce and analyse measured meson and proton
spectra and also make predictions for anti-protons, deltas, anti-deltas
and hyperons.  
The low $m_t$ enhancement in $\pi^-$ spectra is explained by
baryon number conservation and strangeness equilibration. 
We also find that negative kaon data are more sensitive 
to the eos, as well as the $K^-/\pi^-$ ratio. 
All hyperons and deltas
are sensitive to the presence of a phase transition in the
forward rapidity region. $\bar{p}$, $\Omega$ and
heavy anti-baryons are sensitive in the whole rapidity range.
\end{abstract}
 
\newpage

\section{Introduction}

Two main conclusions can be drawn so far from the study
of heavy ion reactions at $AGS$ and $SPS$ accelerators:
\\
(i) nuclear matter is not transparent \cite{stopping}.
In particular for collisions of heavy nuclei at AGS ($Au+Au$) 
the shape of the proton rapidity 
density distribution around the center-of-mass rapidity
suggests an almost total nuclear stopping \cite{prpt},
which also means that high baryon densities are achieved
\cite{QM95}. 
\\
(ii) The assumption of local thermodynamical
equilibrium leads to an astonishing agreement with the data. 
 This follows among other things from the fact that
simple fireball models \cite{regensburg},\cite{fireb2}
which take into account a longitudinal flow component
can explain many features of the data. 

These aspects justify the investigation
of heavy ion physics with more realistic hydrodynamical models
which - if applicable - would serve as a powerful tool for the
description of strongly interacting many particle systems  
and hadronic multiparticle production (\cite{strot}-\cite{dan}).
 
The basic hydrodynamical model
is a generalization of statistical models introduced in the early
fifties \cite{fermi}. 
It was introduced by Pomeranchuck and Landau
\cite{La}-\cite{Bemi} who removed several weak points of the
previous fireball models. The unrealistic concept of a
fireball in global equilibrium, which is not consistent
with the covariant relativistic dynamics of the
collision, was replaced by the concept of a system in
local equilibrium.

  The latter concept is more general and takes into account that the whole 
system is not yet completely equilibrated, but has inhomogeneities
caused by the initial dynamics, which are controled by the strong interaction.
It also takes into account that a system at very high temperatures
does not only evaporate particles from the surface
but also has to expand because
of the strong internal pressure. The details of the expansion
are determined by the equation of state (eos), which describes 
the properties of strongly interacting hot hadronic matter.

The expansion leads to a cooling of 
the system which changes the absolute particle yields,
the chemical composition of the
fireball (particle ratios), the momentum distributions,
as well as the mean free path, which 
increases with decreasing density (or temperature)
of the system. If the mean
free path is large enough the particles decouple (freeze-out)
from the fireball. 

The concept of local equilibrium and relativistic covariance
also requires that 
decoupling takes place locally, i.e. the particles are emitted
when the fluid cell reaches the decoupling temperature $T_f$ 
\footnote{In this work we choose a critical temperature
$T_f$ on the order of the pion mass for the freeze-out
criterion.}.
In other words, below $T_f$ the mean free path becomes 
too large in order to maintain equilibrium. 
A local freeze-out usually leads 
to a very complicated shape of the emission region in space-time
(the freeze-out hypersurface).

From the hydrodynamical point of view 
it is convenient to divide a 
heavy ion collision into 3 stages:
 
\begin{enumerate}
\item The compression and thermalization of nuclear matter forming the
locally equilibrated fireball ({\it compression stage}).
 
\item The hydrodynamical expansion of the fireball ({\it expansion stage}).

\item The decoupling of particles ({\it freeze-out}).
\end{enumerate}

   Supported by the observation of a high amount of stopping, 
we have extended the 3-d hydrodynamical
description to the very beginning of the fireball formation process.
The applicability of the numerical code HYLANDER \cite{ornik},
which exactly solves the relativistic
hydrodynamical equations (Eq.1) and which we used in, e.g. \cite{Bo},
\cite{Bobs},\cite{blei1}, 
is now amplified to simulate the whole collision,
from the very moment the two nuclei touch each other. 

The purpose of this work is: 1) to apply the above
hydrodynamical formalism to $Au+Au$ reactions at $AGS$
in order to investigate the eos as well as the possibility
of a phase transition and 2) to make predictions for yet unobserved 
particle species and their spectra.
\\
\\
The paper is organized as follows. In Sec. 2 we introduce
the equations of state under investigation. 
In Sec. 3 we describe the hydrodynamical model.
In Sec. 4 the results of the simulations are shown. 
It contains an analysis of the equations
of state and the corresponding properties of the 
simulated fireballs,
and the comparison
with published data for protons, pions and kaons, 
as well as predictions
for anti-protons, heavy baryons and heavy anti-baryons.
In Sec. 5 we present a description for the 
negative pion enhancement and discuss strange particle rates. 
Sec. 6 contains a discussion of our results.

\section{Hydrodynamical Model}

In their
simplest form the hydrodynamical equations
do not include dissipative effects. 
The incorporation of dissipation in a relativistically
covariant way is up to now very difficult 
and requires approximations. Some progress in this
field has been made for two and three fluid dynamics
and dissipative shock waves \cite{mornas}-\cite{galipo}.
In the following we will restrict ourselves to one fluid described by the 
relativistic Euler equations,

\ba
\frac{\partial E}{\partial t}&=&-\nabla\left((E+P){\bf v}\right)
\qquad\mbox{\hskip0.1cm energy conservation} \nonumber\\
\frac{\partial M^i}{\partial t}&=&-\nabla(M^i{\bf v})-
\frac{\partial P}{\partial x_i}
\qquad\mbox{momentum conservation}\nonumber\\
\frac{\partial (b\gamma)}{\partial t}&=&-\nabla b {\bf v} \qquad
\mbox{\hskip1.8cm baryon number conservation}
\nonumber\\
E&=&\gamma^2(\varepsilon+P v^2) \nonumber\\
{\bf M}&=&\gamma^2(\varepsilon+P){\bf v} \nonumber\\
\ea

where ${\bf v}(\vec x ,t)$ is the velocity of the fluid, 
$\gamma= 1/\sqrt{1-v^2}$,  $P$ is the pressure,  $\varepsilon$ 
is the energy density and $b$ is the baryon density.
\\

   The solution of these equations is determined by the equation of state
which can be written in the form
\be
P=P(\varepsilon,\mu),
\ee

where $\mu$ is the chemical potential.
It governs the compression, the expansion and the
freeze-out surface shape of the fireball.

If the local density drops below a critical value ($\rho_f$)
the particles are assumed to decouple (locally) 
from the fluid, i. e. hydrodynamics is not applicable
beyond this point.
The primordial resulting particle
spectra are then described by the Cooper-Frye formula
\cite{cfrye}:

\begin{equation}
E\frac{dN}{d\vec p}=\frac{g_i}{(2\pi)^3} \int_\sigma  
\frac{P_{\mu} d\sigma^{\mu}}{exp(\frac{P_{\mu} u^{\mu} - \mu_s-\mu_b}
{T_f}) \pm 1},
\end{equation}
\\
which describes the distribution of particles with degeneration
factor $g_i$ and 4-momentum $P^{\mu}$ emitted from a 
hypersurface element $d\sigma^\mu$ with 4-velocity $u^{\mu}$.
After the cascading
of the resonances we obtain the final observable spectra.

   In our earlier approach \cite{Bo} at $SPS$ energies we took into account
that due to transparency effects the local equilibrium state 
is reached after undergoing a non-equilibrium stage, which 
is not treatable with hydrodynamics.
Therefore we started our simulation in an intermediate state 
which we had to model by introducing some parameters 
based on ``reasonable" assumptions about the initial configuration.
However both for $S+S$ and $Pb+Pb$ \cite{blei1} collisions
we found that the inelasticity necessary to describe the data
was larger than $70 \%$.
For lower energies ($AGS$) the inelasticity (or
amount of stopping) is expected to increase. 

Therefore in the present paper
we extend the HYLANDER-code using
a different approach to the hydrodynamics of
heavy ion collisions suited for processes with (almost)
full-stopping. It is based on the original Landau model
\cite{La},\cite{La88} where the process of stopping is also
treated hydrodynamically, rather than being parametrized
by some initial conditions as in \cite{Bo},\cite{Bobs}.
As will be explained below,
our approach constitutes an important improvement
over the old Landau approach as it eliminates
all approximations he made.

The starting point of the model are two colliding tubes at 
zero temperature and nuclear ground state density. 
The width in longitudinal (beam) direction is
given by the Lorentz-contracted
nuclear diameter in the equal velocity frame.
This problem was solved by Landau analytically \cite {La} 
with the following approximations:

\begin{enumerate}
\item  the assumption of 1-dimensional shock waves, 
\item  the 1-d hydrodynamical description for the beginning
of the expansion process followed by an approximated 3-d 
analytical solution and
\item  an equation of state of the type $P=c_0^2  \varepsilon$ 
where $c_0$ is the constant velocity of sound.

\end{enumerate}

   In the present work we
do not use any of these approximations, since 
both the {\it compression stage} and the {\it expansion 
stage} are described by a fully
3-dimensional hydrodynamical simulation. 

As a consequence we have additional contributions to
the particle spectra from the very early {\it compression stage}.
The $Au+Au$ system at $AGS$ spends about $4$ $fm/c$ in 
this stage\footnote{ Landau 
made an $1-d$ approximation
of these stage and neglected its contribution to the spectra.
Such a method is only justified at extremely high energies
where the Lorentz contracted longitudinal diameters before
collision are very small compared to the lifetime of the system,
which is not the case for $AGS$ energies.
Squeeze-out effects and transverse motion are also
not present in Landau's approach.}. 

Given the fact that, due to this treatment, there are no free parameters
necessary to describe the initial conditions of the fireball,
it becomes now possible to study the sensitivity
of the results to the properties of the eos. This will
be done by applying our solutions of 3-d hydrodynamics
to $Au+Au$ reaction at the $AGS$.

\section{The Equations of State}

In this work we considered two models for the eos
as an input to solve numerically the relativistic
hydrodynamic equations with the HYLANDER code:
\\

1) Firstly we present an 
eos given by a parametrization \cite {latp}
of lattice-QCD results \cite {lat}. 
It describes a first order
phase transition
between quark gluon plasma and hadronic matter at $T = 200 MeV$
and corresponds to a baryon chemical potential $\mu = 0$
\footnote{In \cite{lat} a pure gluonic system is considered.
Results considering dynamical quarks lead to a critical temperature
$T_c$ between $150$ and $200 MeV$ \cite{lat2}.}.
The baryons are explicitly
considered in the hydrodynamical equations.
We will refer to this eos as {\bf lattice-eos}.
\\

2) Secondly we took two versions of a resonance gas
equation of state \cite {mornas},\cite {resg}, 
which differ in the number of included resonances. 
In the following we will
refer to {\bf RG1.5} for a resonance gas including resonances
with masses up to $1.5$ $GeV$ and {\bf RG2} for a gas 
of resonances of masses up to $2$ $GeV$. In this eos 
the dependence on the baryon chemical potential and strangeness
conservation is included.

\section{Results}

\subsection{ Energy density, baryon density and lifetime}

Table 1 shows the values for 
maximum energy density, maximum baryon density and 
the time it takes until
the fireball is completely transformed into 
free particles (lifetime). All these simulations start 
at the moment of the impact between the nuclei ($t=0$).

The system spends one third of its lifetime in the
compression stage, confirming the importance of
this part of the process.

The values for lifetime, maximum baryon and energy density for
lattice-eos and RG2 are surprinsingly similar, but this does
not necessarily mean that the behaviour of each fluid cell
(and of the whole fluid) until freeze-out is also the same.
To investigate this we study the trajectory of the fluid elements 
in a phase diagram of energy density versus temperature, for 
the three eos (see Fig.1). In this figure we plot
the temperature and energy density 
for each fluid cell
with $T > 139$ $MeV$ (which means for all $\vec x$ and $t$),
starting at the beginning of the simulation.

One sees that for lattice
and RG2 the fluid elements describe a trajectory
in almost the same energy density and temperature range. 
For both simulations the fluid elements can reach temperatures
up to $215$ $MeV$ and energy densities bigger than $6$ $GeV/fm^3$,
for lattice-eos this correspond to 
temperatures slightly above the phase transition temperature.

The trajectory for RG1.5 is located in a very different range.
The explicit dependence on a baryonic chemical potential in RG2
and RG1.5 appears just in the ``width" of the curve.
It is interesting to note that even in the case
of RG1.5 and RG2 where $\mu$ enters explicitly in the calculation
the baryon dependence is weak, i.e.
the ``width" of these curves is surprisingly
small.

The similarity in the results for energy density, baryon density and
lifetime for lattice-eos and RG2 
can be explained by Hagedorn's model \cite{resg}. 
Increasing the number of resonances in the hadronic gas eos
induces a phase transition-like behavior in the 
development of the fireball
\footnote{An important and obvious question is whether
this behaviour 
of the fluid elements will
be the same for other nuclear reactions. 
We performed a simulation for
$Pb+Pb$ at $160$ $GeV/nucleon$ to try to answer this question. 
We used the same initial condition as
in the present paper and two eos, namely lattice and RG2. 
The resulting trajectory of the fluid elements 
differs from that for
the $AGS$ system. The difference between both eos 
appears for temperatures larger then $0.2$ $GeV$.
What this implies for the particle spectra
will be discussed elsewhere \cite {nblei}.}.

The sensitivity of the produced particle spectra to these 
differences in the eos is the subject of the following sub-sections.

\subsection{Particle spectra}

At freeze-out temperature
we treat explicitly the emission of
protons, neutrons, pions, kaons, anti-protons
(directly produced) 
and the particles/resonances:
$\Omega(783)$,$\eta$,$\eta\prime$,
$\rho$,$K_0$,$K^*$,
$\Delta$,$\Sigma$,$\Lambda$,$\Xi$ 
and correspondent anti-particles (see \cite{Bo}).
The results we present in the next sub-section
take into account the contribution from
the decay of particles.

\subsubsection{Spectra of protons, pions and kaons}

In Figure 2 we compare the transverse mass 
($m_t=\sqrt{m^2+p_t^2}$)
spectra of protons,
positive and negative pions for different rapidity intervals
from our {\bf RG1.5} simulation with experimental 
data \cite{prpt},\cite{QM95}. 
All spectra obtained for this eos differ considerably
from the data.

In Figure 3 we show the corresponding results 
from our simulation using the {\bf lattice-eos}.
One can see that, except for the very central region (last curve),
where the proton production is overestimated,
the experimental proton spectra are very well fitted.
The fits for positive and negative pions 
show the same tendency: significant deviation from
the data we observed only in the very central rapidity
region (first curve).
In all cases there is a small overestimate of particle production
at large $p_t$, which means in hydrodynamical terms
an overprediction of transverse flow.

In the last section we mentioned the similarity in
the lifetime, baryon and
energy density, as well as the trajectory of the fluid elements
arising from lattice-eos and RG2 simulations.
Therefore we expect that the RG2 spectra
are more similar to the spectra obtained using lattice-eos
than to the ones using RG1.5.

The simulation using RG2 confirmed this expectation. 
One can see this
in Figure 4 where we consider 
the $m_t$ spectra resulting from the
simulation with {\bf RG2}. 
Particularly for pions we observe a very good agreement
with the data. Even the overestimate of pions at large
$p_t$ observed for the lattice-eos simulation vanishes.
In the case of protons we observe deviations only
in the very central rapidity region.

In Figure 5 we compare the spectra, at fixed rapidity, 
for all three simulations. 
Here one can see that for the protons 
the best fit is clearly given by the lattice-eos simulation.
For positive pions the simulation by RG2 is better and
for negative pions the best results are obtained with both
lattice-eos and RG2.

If we take into account the so far available data
we can already conclude that a medium with
a large number of
internal degrees of freedom (a very ``soft" eos) 
is favoured.

All three simulations show at midrapidity 
an overestimate of proton and pion yields.
We attribute this effect to the transverse squeeze-out of nuclear matter
which has its maximum at zero impact parameter (central collisions). 
We recall that the data are sampled over a finite
impact parameter ($b$) region 
\footnote { The ``very central" data ($4\%$ centrality)
for $Au+Au$ $AGS$ correspond to an impact parameter $b<$$2.6$ $fm$ 
(Y.Akiba, private communication and QM96).}
, whereas the simulation is really at $b=0$,
therefore the squeeze-out appears diluted in the data.
 
However we do not encounter the problem cited in
\cite{prpt} (and references quoted there for Monte Carlo models)
which could not reproduce the flatness and shape of the proton and
pion spectra.

The enhancement at low $m_t$ exhibited in the $\pi^-$ spectra 
is present in all three simulations. We can reproduce this effect
in a natural way just by taking into account
resonances, baryon conservation and strangeness equilibration,
without invoking statistical
and systematic errors in the data, as was done in \cite{QM95}.
This manifests itself not only as a change in the shape
of the $\pi^-$ spectra but also as an increase of the total multiplicity
of negative pions compared to the positive ones.
A more precise analysis of $\pi^-$ and $\pi^+$ production
will be presented in section 4.4. 

Now we turn to rapidity 
distributions \footnote{ The rapidity distributions from our
model do not contain the phase space cuts at low $m_t$
which are present in the data.} (Fig. 6). 

For protons and pions 
the already observed tendency is confirmed, namely the
results arising from the simulations with
lattice-eos and RG2 are closer to the data than RG1.5. 

In pion production one can see here explicitly
that the hydrodynamical simulation
produces more negative pions than positive ones, 
a fact which is confirmed by
the experiment.

In the rapidity distribution analysis we
include a comparison
between our results and strange particle production data.
This is of particular importance because of the well known
proposal to look at strangeness production as a signature of QGP
(cf. e.g. \cite{stock} - \cite{lee} for more recent references).

In Figure 6 one also can see that
for the kaon rapidity spectra the difference between 
the three simulations is more pronunced,
particularly for negative kaons.
The comparison with preliminary
data favours the lattice-eos.
\\

As an preliminary conclusion for this sub-section
we can say that the results are generally in surprisingly
good agreement with the 
data, especially if one 
takes into account that we do not need
any parameters other than those that enter the eos.

We also see that
the presence or absence of a phase transition 
can not be determined by the analysis of $m_t$
spectra of protons and pions.

The situation appears to be different if we look at other aspects such as
the total number of produced protons and pions (see Table 2) and
the rapidity distributions of protons, pions and especially 
kaons, where we observe remarkable differences
between the spectra resulting from the three simulations. 

Since kaon production in a baryon-rich medium
is linked to hyperon production and chemical equilibration
we expect from those also a sensitivity in the hyperon yield
related to the eos.
Motivated by this fact we will investigate in the following
the rapidity distributions of deltas, hyperons
and their corresponding
anti-particles. We also consider anti-proton
production.

The data are generally better described
if one uses a ``softer" eos. Because of that 
from now on we will 
restrict the discussion to the results from the simulations
with the lattice-eos and RG2.

\subsubsection{Predictions for anti-baryons and heavy baryon production}

In Figures 7,8 and 9 we show the rapidity distributions for 
anti-protons, heavy baryon and heavy anti-baryon production. 
Table 2 shows the total particle number for all created particles
and anti-particles for the three simulations.

The lattice-eos produces a larger or a comparable 
number of heavy baryons and heavy anti-baryons than RG2
(except for $\Delta$).
The largest differences are predicted for $\Xi$, $\Omega$, 
$\bar{\Delta}$
and $\bar{p}$ \footnote{The direct $\bar{p}$ production
contributes with only $\sim 30\%$ to the total $\bar{p}$ abundancy
total numbers.}.

For heavy baryons (Fig. 8) differences appear in the forward rapidity
spectra ($y > 1.0$). The $\Omega$ production differs in the whole 
rapidity range.
For the heavy anti-baryons rapidity spectra (Fig.9) 
differences appear in the whole rapidity range,
except for $\bar{\Xi}$, where the difference appears
only in the forward region $(y > 1.0)$.

\section{Highlights} 

In the following we would like to emphasize three remarkable results of our 
calculations:
$(a)$ the enhanced $\pi^-$ production compared with $\pi^+$ production;
$(b)$ the ratio $N_{\Xi^-}/N_{\Lambda}$; and c) the rate
$K/\pi$.
\\
\\
$a)$ Taking into account baryon and strangeness conservation
as well as strangeness equilibration at $AGS$ energies
and including the decay of the resonances 
in the final stage in our model the difference $N_{\pi^ -} - N_{\pi^+}$ 
is given by

$N_{\pi -} - N_{\pi +} = 
0.64 N_{\Lambda} + N_{\Sigma ^-} + N_{\Xi -} + 0.64 N_{\Xi 0}$,\\
i.e., the hyperons and their decay mainly determine
the difference between $\pi^-$ and $\pi^+$ total numbers.
Since our model fits both the $\pi^-$ and $\pi^+$
spectra, we have a natural and simple explanation
for the experimentally observed difference in the multiplicities
of positive and negative pions.
This also explains the experimental observation from flow analysis 
that $\pi^-$ flow is correlated with the 
protons \footnote{T.Hemmick private communication and QM96.},
as the $\pi^-$ contribution from hyperon decays is connected
to the baryon flow\footnote{
In reference \cite{coulomb} the authors investigate
the low $m_t$ enhancement in $\pi^-$ production for
$Pb+Pb$ system at $SPS$ energies and attribute it to the Coulomb
effect. It is likely that the explanation presented above for the
$AGS$ data applies in this case too.
Results on this subject will be presented elsewhere. This sugests
that the Coulomb effect invoked to explain these data is probably
much weaker than assumed in \cite{QM95}\cite{coulomb}.}.
\\
\\
$b)$ In \cite{hyp} multi-strange hyperons ($S \ge 2$) 
and strangelets are suggested as better signatures than single-strange
particles ($S = 1$). The first report about $\Xi^-$ 
production in heavy ion collisions 
at AGS \cite {chi} (for $Si + Pb$) mentions that the observed
rates are at least 5 time bigger than all present cascade model
predictions.
There are no such data avaiable for $Au+Au$ collisions at $AGS$ energies
at the moment,
but $Si+Pb$ at $AGS$ energies constitutes an experiment
in the same energy range, with a high baryon density
and a high amount of stopping. Encouraged by these aspects we compare
our results with these data.
  
For $Au+Au$ we find the following results:

$N_{\Xi^-}/ N_{\Lambda} = 0.126$ (lattice-eos)

$N_{\Xi^-}/ N_{\Lambda} = 0.090$ (RG2)

which are in agreement
with the experimental value of $0.12 \pm 0.02$
for $Si + Pb (AGS)$.
However the difference between lattice-eos and RG2 simulations
is not very big which leads us to the 
tentative conclusion that the $\Xi^-$ production does not 
necessarily serve as a better signal for
QGP than other strange particle yields.
\\
\\
$c)$ The experimental ratios $K/\pi$ for $Si+Au$ $(AGS)$
were measured  and published in \cite{dkp}. The values are 
$K^+/\pi^+ = 0.192 \pm 0.03$ and $K^-/\pi^- = 0.036 \pm 0.008$ in
the mid-rapidity region and they have been presented
as intriguing results because of the large strangeness yields
compared with $S+S$ $(SPS)$, where both values are
about $0.11$.
For $Au+Au$ $(AGS)$ experiment the ratio $K^+/\pi^+$  
is found to be 0.21
\cite{QM95} and the negative ratio was not yet published.

The results are remarkable in the sense that 
there are no strange 
particles present in the initial state and a significant 
rate of strange particle production is only understandable 
if a strange chemical equilibrium is established during 
the reaction. Strangeness equilibration however is not easy 
to justify in a pure hadronic scenario.

{}From our calculations, which includes the assumption 
of strangeness equilibrium, we find:

$N_{K^+}/N_{\pi^+} = 0.207$ (lattice-eos) and $0.256$ (RG2), \\
which are in 
surprisingly good agreement with the experimental results, 
especially for the lattice-eos scenario. 
The numbers from the lattice-eos and RG2 simulation
are not very different and we can conclude therefore that this 
result is a strong indication
of local equilibration (including strangeness equilibrium) 
of the system. The signal however is not very sensitive
to the concrete type of the eos.

On the other side, clear differences appear (see Table 2) for
the other ratio:
$N_{K^-}/N_{\pi^-} = 0.038$ (lattice-eos) and $0.018$ (RG2).
Experimental information about this ratio is very important
and should be treated with care.

{}From our results we can conclude that
a hadronic scenario considering strangeness equilibration
can also explain the difference between the positive 
and the negative ratios \footnote{Other aspects in this discussion 
related to $Si+Au$ $(AGS)$ is presented in \cite{cleymans}}.

\section{Conclusions}

We have demonstrated that the 3d-hydrodynamical model 
presented above can describe quite reasonably the
$AGS$ data for $Au+Au$ reactions. This suggests
that the hypothesis of local thermodynamical equilibrium
applies also for the early stage of the reaction.

We showed that 
both an eos based on QCD lattice calculations 
exhibiting a phase transition between quark gluon plasma/hadronic
phase (lattice-eos) and a resonance gas eos including 
resonances with masses up to 2 GeV (RG2) 
have the essential physical properties necessary to describe the
mesured proton and pion $m_t$ spectra.
An eos described by a resonance gas with a 
small number of degrees of freedom (RG1.5) 
is not consistent with these data. 
However as shown by Hagedorn \cite{resg} 
the RG2 scenario is 
related to the idea of a phase transition.
This phase transition-like behaviour becomes even
more pronounced if one adds higher resonances.
However, we note that the
assumption of strange chemical equilibration , which is assumed to 
be  present even in this hadronic scenario, is not easy to 
justify in the case of a pure hadronic eos.

In a general analysis including the $m_t$ spectra for 
protons and pions, the total multiplicities 
of produced protons and pions and their
rapidity distributions we can conclude
that the lattice-eos provides an overall better description of
the $Au+Au$ $(AGS)$ experimental data than a hadronic eos.

We have also calculated the particle spectra 
for anti-baryons and heavy anti-baryons 
and the rates $\pi/K$ and $\Xi^-/\Lambda$ 
in order to investigate the influence of a phase transition 
on the production of these particle species.

Generally the simulation with the equation of state containing a 
QGP-hadronic phase transition between a hadronic phase and a QGP 
predicts a larger total multiplicity 
of heavy baryons and anti-baryons than with resonance gas.  
The largest differences in the number of produced
particles appear for $\Xi$, $\Omega$, $\bar{p}$ and 
$\bar{\Delta}$.

In all heavy baryon and heavy anti-baryon rapidity distributions,
the strong difference between both eos scenarios appears 
in the forward region of rapidity, $y > 1.0$.
For $\Omega$ and heavy anti-baryons (except for $\bar{\Xi}$)
differences are also predicted in the mid-rapidity
interval. Despite the
lower multiplicities this could be an interesting topic
for future experiments. 

Negative kaons we found to be more sensitive to the presence of
a phase transition in the eos than the positives ones, 
as well as the corresponding
$K^-/\pi^-$ ratio. The ratio $K^+/\pi^+$ has been compared with
experimental data and and both scenarios (specially the lattice 
one) are in good agreement with the measured ratios,
a fact which again supports the assumption of an almost 
complete chemical equilibration.
Differences between the positive and the negative ratios
were found in both scenarios.

The rate $\Xi^-/\Lambda$ was compared 
with experimental data (for $Si+Pb$)
and is in good with them.
The results for both scenarios are not very different
and therefore we conclude this ratio involving a 
multi-strange hyperon
does not appear to be a better signature then
the $S = 1$ particle yields. 

A particularly important aspect of our investigation is that
the high negative pion multiplicity in this experiment can be obtained
in a natural way just taking into account 
baryon and strangeness conservation, strangeness equilibration
and resonance decays. 
It has its origin mainly in
the  $\Lambda, \Sigma$ and $\Xi$ channels, 
as we showed in detail in the previous section.
The low $m_t$ enhancement in $\pi^-$ spectra
can also be explained in this way. 
We conclude that, contrary to the statement made in
\cite{coulomb}, the Coulomb effect does not strongly
affect the pion spectra.

Another step in the investigation of the equation
of state which governs the heavy ions physics would
be to realize the same simulation
using an eos based on lattice-QCD calculations 
extended into the baryonic sector.
\\
\\
\\
N.A. wants to thank T. Hemmick and
Y. Akiba for important discussions.
We would like to thank J. Bolz, D. Strottman and B. Schlei for 
computational help. 
R.W. is indebted to A. Capella
for the hospitality extended at LPTHE, Univ. Paris-Sud.
U. Ornik thanks GSI Darmstadt and Soultek Internet Services
for financial and computational support. 
This work was supported in part by the CNPq-Brazil (Brasilia).

\newpage

{\huge\bf Figure Captions}

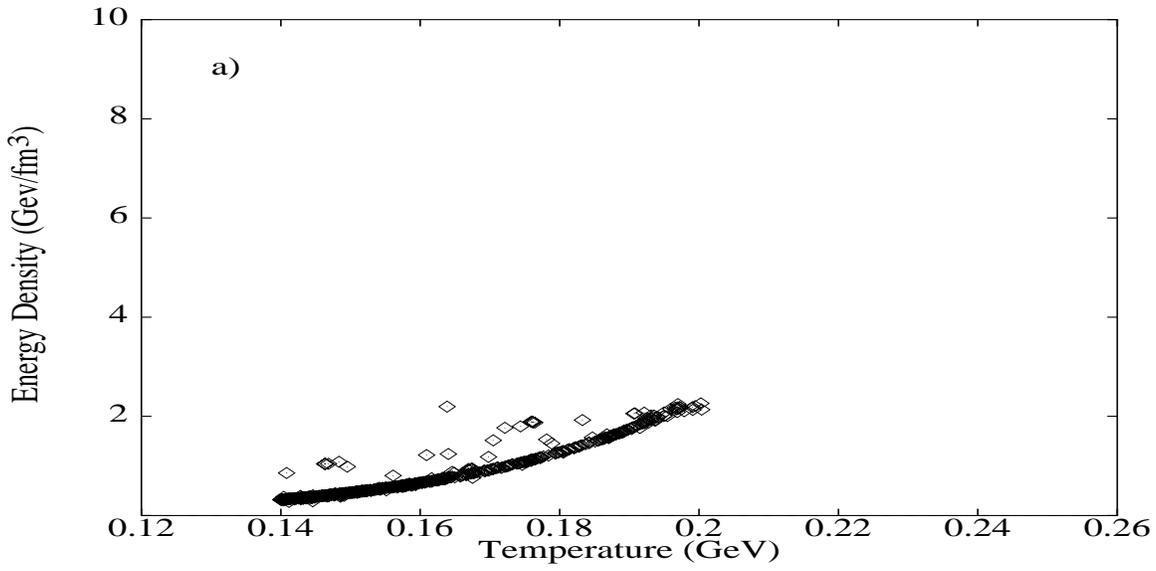
{\bf Fig. 1:} Plot of temperature and energy density of
each fluid cell with $T > 139$ $GeV$, from the beginning of
the collision, for the hydrodynamical simulation 
(HYLANDER) using
the different eos. The figure shows
the trajectory of the cells in the $(\epsilon,T)$
diagram until they freeze out.
a)for RG1.5, b) for lattice-eos and c) for RG2.
 
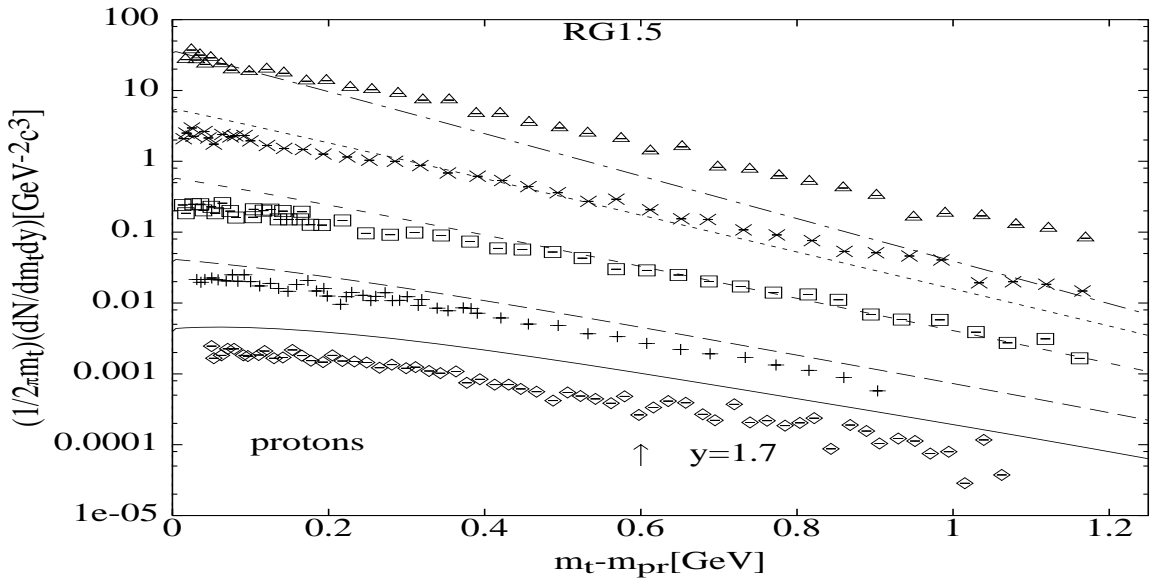
{\bf Fig. 2:} Transverse mass spectra for five rapidity intervals
for protons, positive and negative pions using
the equation of state 
RG1.5. The data (taken from \cite {QM95} \cite {prpt})
and hydrodynamical simulated curves obtained for hydrodynamical
simulation (HYLANDER) 
are shown for rapidity bins from $1.7$ to $2.5$ (for pions),
from $0.9$ to $1.7$ (for protons). In both cases the
bin size is $0.2$ and the bins are centred around $y_{central} = 1.6$.
 
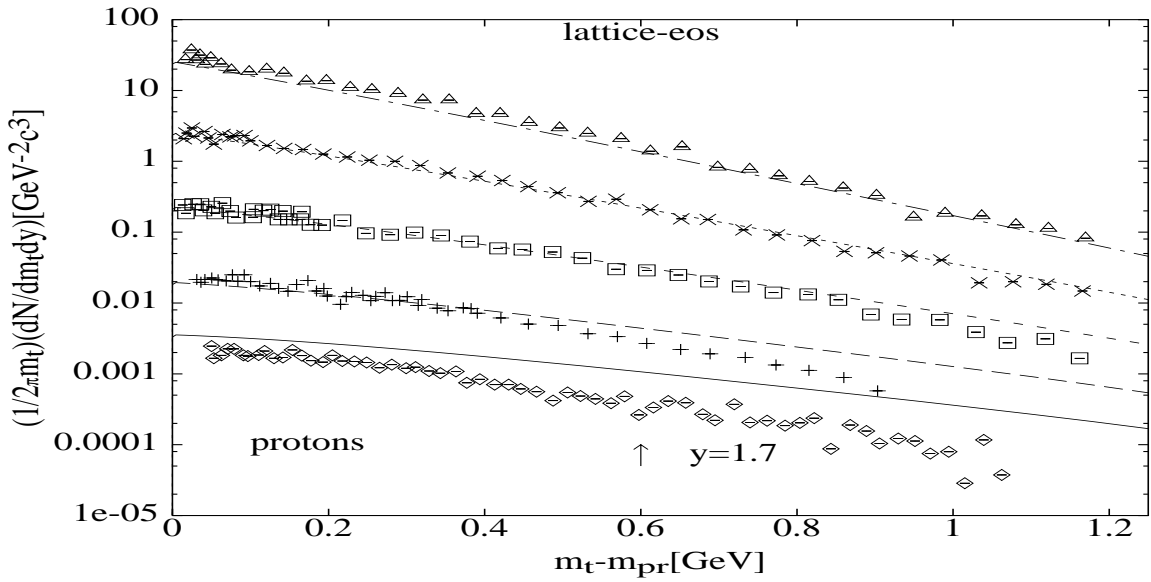
{\bf Fig. 3:} Transverse mass spectra for five rapidity intervals
for protons, positive and negative pions using lattice eos. 
The data and rapidity intervals are the same as in Fig. 2.

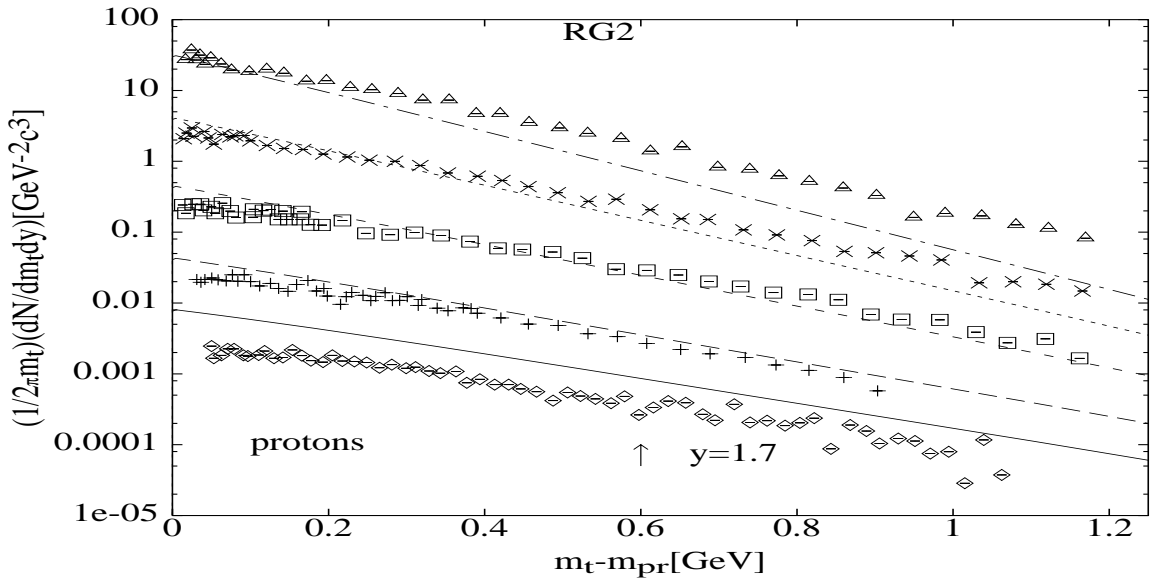
{\bf Fig. 4:} Transverse mass spectra for five rapidity intervals
for protons, positive and negative pions using RG2. 
The rapidity intervals and data are the same as in Fig. 2 and 3.

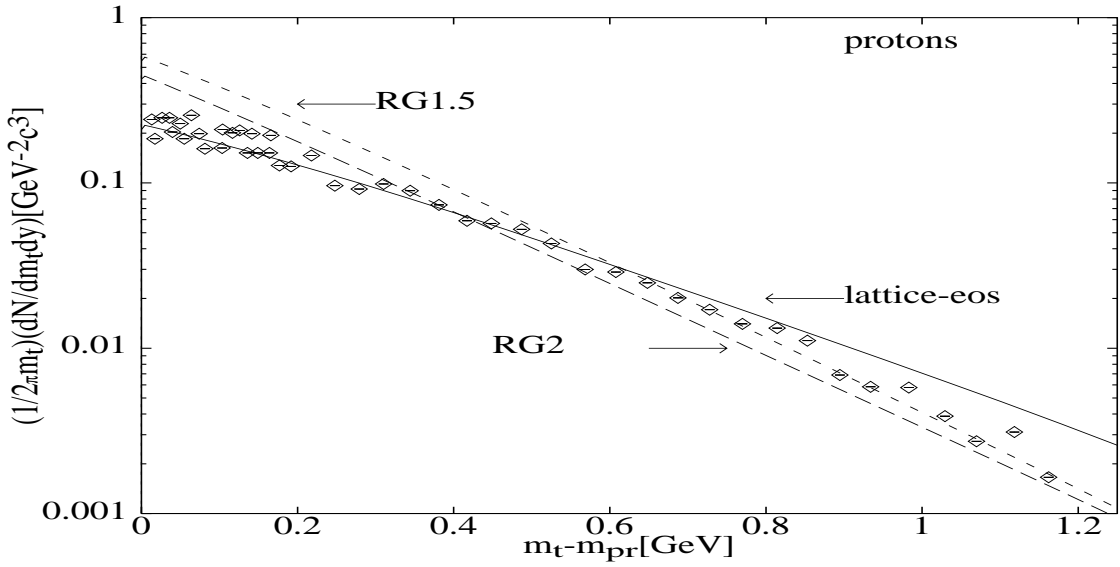
{\bf Fig. 5:} Comparison of the $m_t$ spectra at fixed rapidity 
($y=2.1$ for pions and $y=1.3$ for protons) 
for hydrodynamical simulations (HYLANDER)
using lattice, RG2 and RG1.5 equations of state.
 
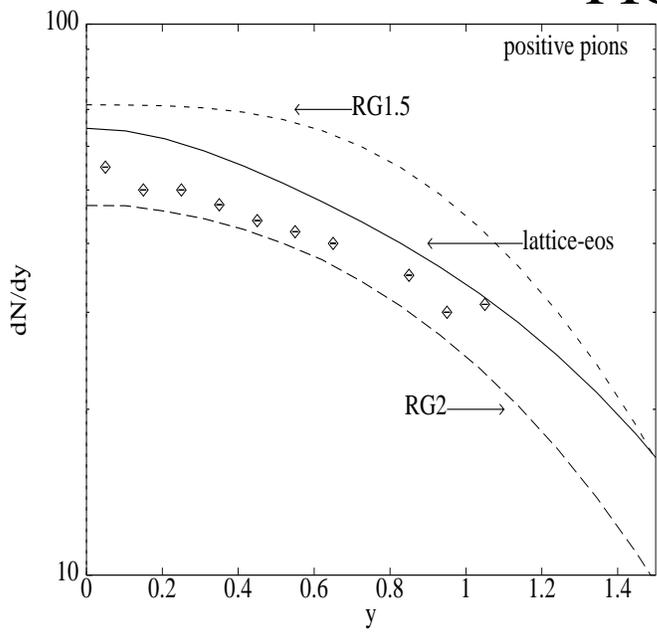
{\bf Fig. 6:} Rapitity distribution for protons, pions and kaons
for hydrodynamical simulations (HYLANDER)
using lattice, RG2 and RG1.5. equations of state.
The data are from \cite {QM95}.
 
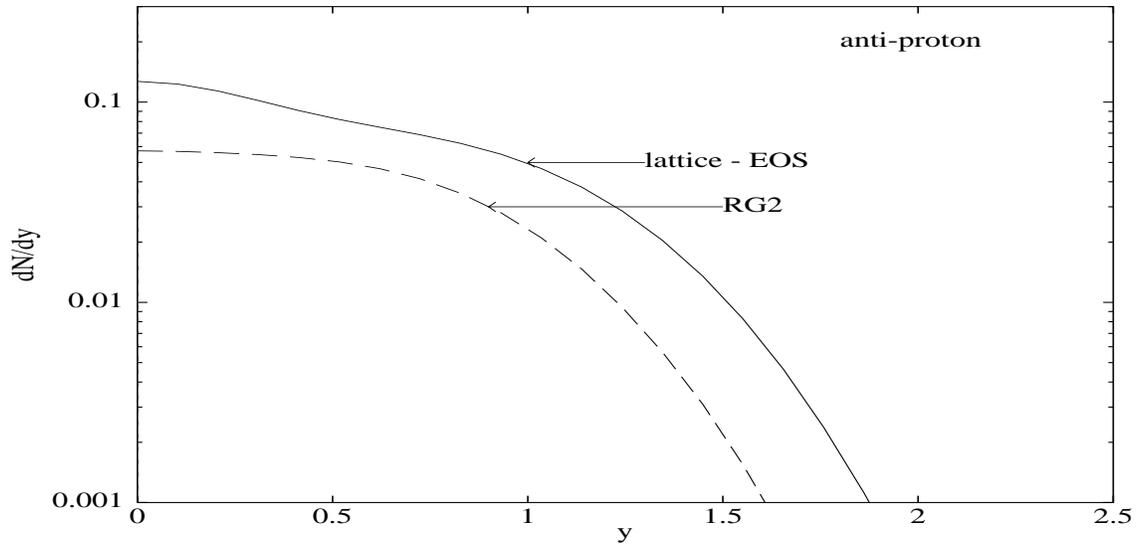
{\bf Fig. 7:} Rapidity distribution for
anti-protons for hydrodynamical simulations (HYLANDER)
using lattice and RG2.

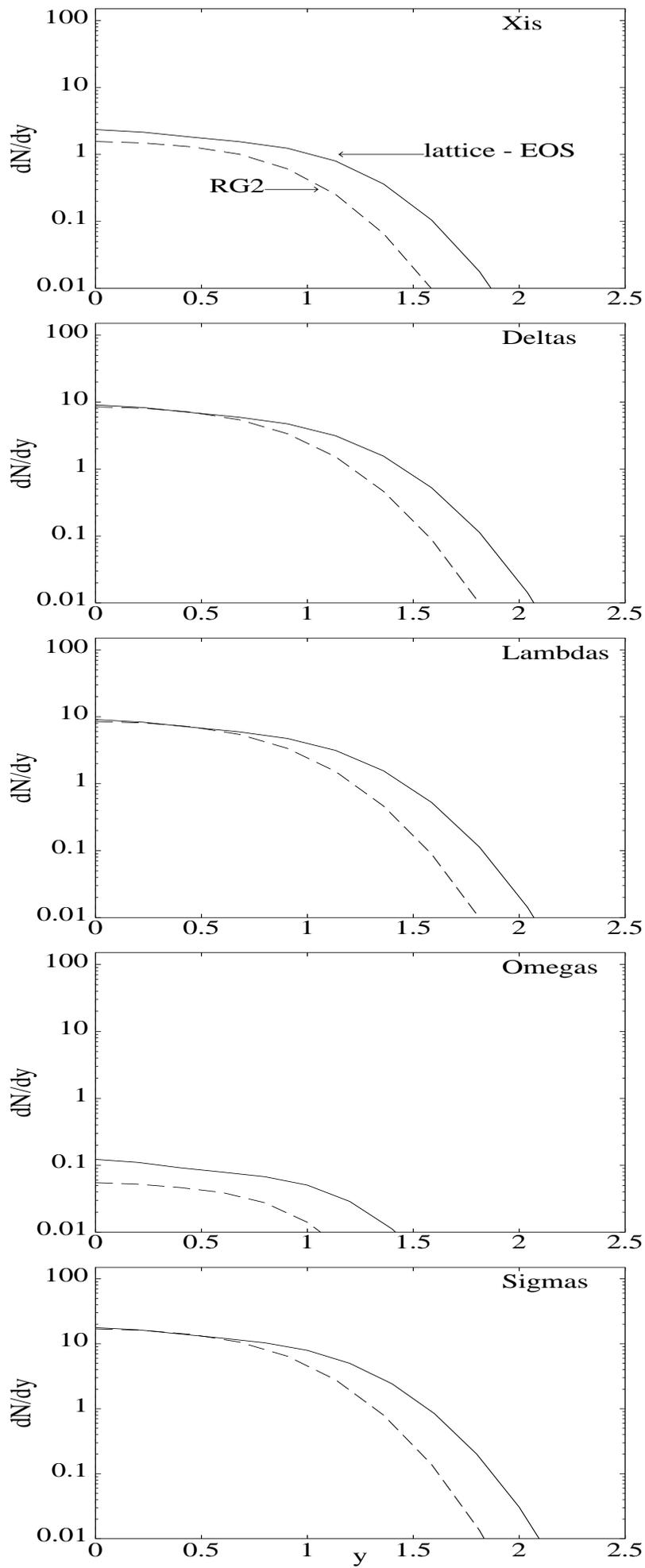
{\bf Fig. 8:} Rapidity distribution for heavy baryon production
$(\Delta,\Sigma,\Lambda, \Omega$ and $\Xi)$ for 
hydrodynamical simulations (HYLANDER) using
lattice and RG2.

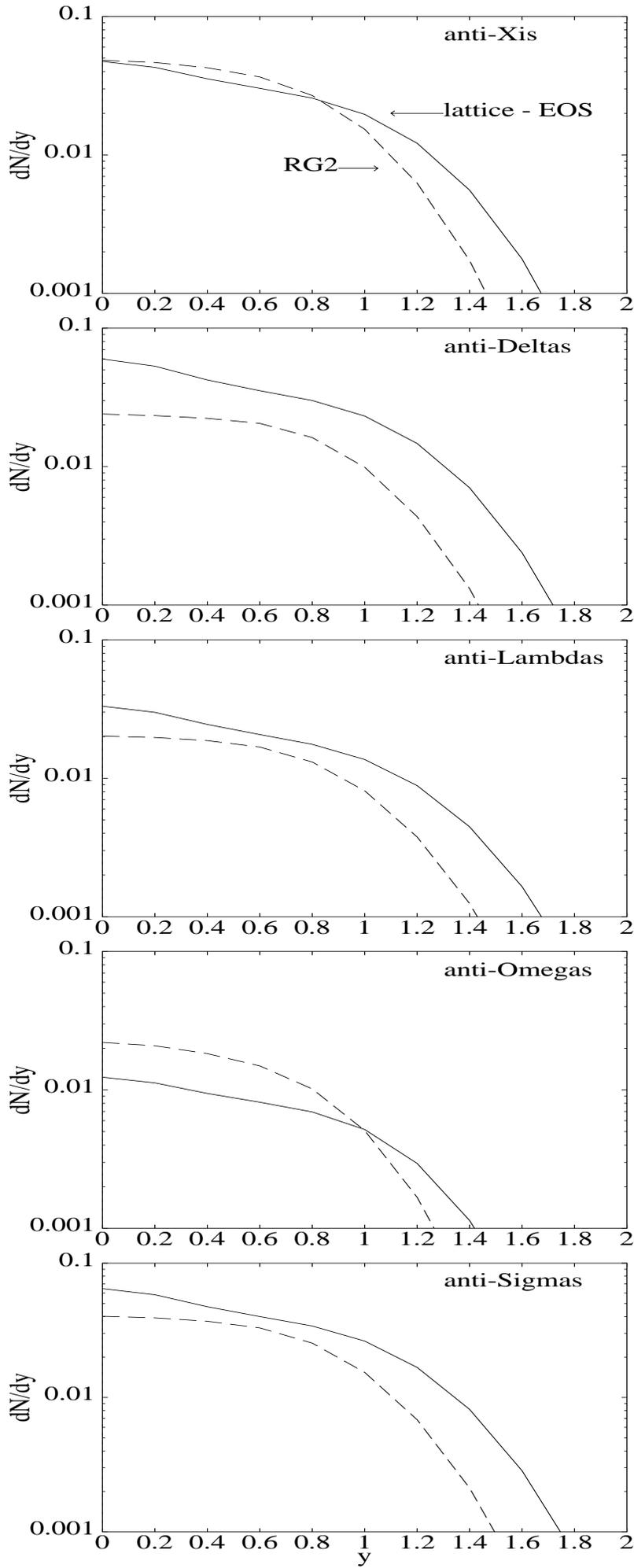
{\bf Fig. 9:} Rapidity distribution for anti-baryons production
$(\bar{\Delta}, \bar{\Sigma}, \bar{\Lambda}, \bar{\Omega}$ and 
$\bar{\Xi})$ for hydrodynamical simulations (HYLANDER)
using lattice and RG2.

\newpage

{\huge\bf Table Captions}

{\bf Table 1:} Maximum values for energy density, baryon density
and lifetime for  hydrodynamical simulations
(HYLANDER) using three different eos. 
$n_0$ is the normal baryon density.

{\bf Table 2:} Total multiplicities of produced particles 
and anti-particles from
simulations using the three different eos. The numbers
take into account the shown isospin degeneracy factors.
The data are from \cite{QM95} (The error bars are around
$10 \%$. Y.Akiba, private communication) .

\newpage

\begin{center}
{\huge \bf Table 1}
\end{center}

\begin{center}
\begin{tabular}{|l||c|c|c|}\hline
  & lattice eos & RG2 & RG1.5 \\ \hline\hline
max. energy density&$6.6$ $GeV/fm^3$&$7.5$ $GeV/fm^3$&$2.5$ $GeV/fm^3$\\ \hline
max. baryon density&$13.6$ $n_0$&$16.7$ $n_0$&$5.6$ $n_0$\\ \hline
lifetime& $10$ $fm/c$ & $10$ $fm/c$ & $15$ $fm/c$\\ \hline
\end{tabular}
\end{center}

\newpage

\begin{center}
{\huge\bf Table 2}
\end{center}

\begin{center}
\begin{tabular}{|r||c|c|c|c|c|}\hline
particle & lattice eos & RG2 & deg. factor & data\\ \hline\hline
p & 132.800 & 140.200 & 1& 160\\ \hline
$\pi^+ $& 140.200 & 100.800 & 1 & 115\\ \hline
$\pi^- $ & 155.000 & 109.000 & 1 & 160\\ \hline
$K^+ $  & 29.000 & 25.800 & 1 & - \\ \hline
$K^- $ & 6.000 & 2.000 & 1 & -\\ \hline\hline
$\Xi$ & 4.100 & 2.400 & 2 & -\\ \hline
$\Delta$ & 153.300 & 197.600 & 4& - \\ \hline
$\Lambda$ & 16.200 & 13.700 & 1 & -\\ \hline
$\Omega$ & 0.200 & 0.080 &1 & -\\ \hline
$\Sigma$ & 31.300 & 26.800 & 3& -\\ \hline\hline
$\bar{p}$ & 0.200 &0.100 & 1& -\\ \hline\hline
$\bar{\Xi}$ & 0.075 & 0.080 & 2 & -\\ \hline
$\bar{\Delta}$ & 0.095 & 0.044 & 4 & -\\ \hline
$\bar{\Lambda}$ & 0.055 & 0.036 & 1& -\\ \hline
$\bar{\Omega}$ & 0.020 & 0.030 &1& - \\ \hline
$\bar{\Sigma}$ & 0.100 & 0.070 & 3& - \\ \hline\hline
\end{tabular}
\end{center}

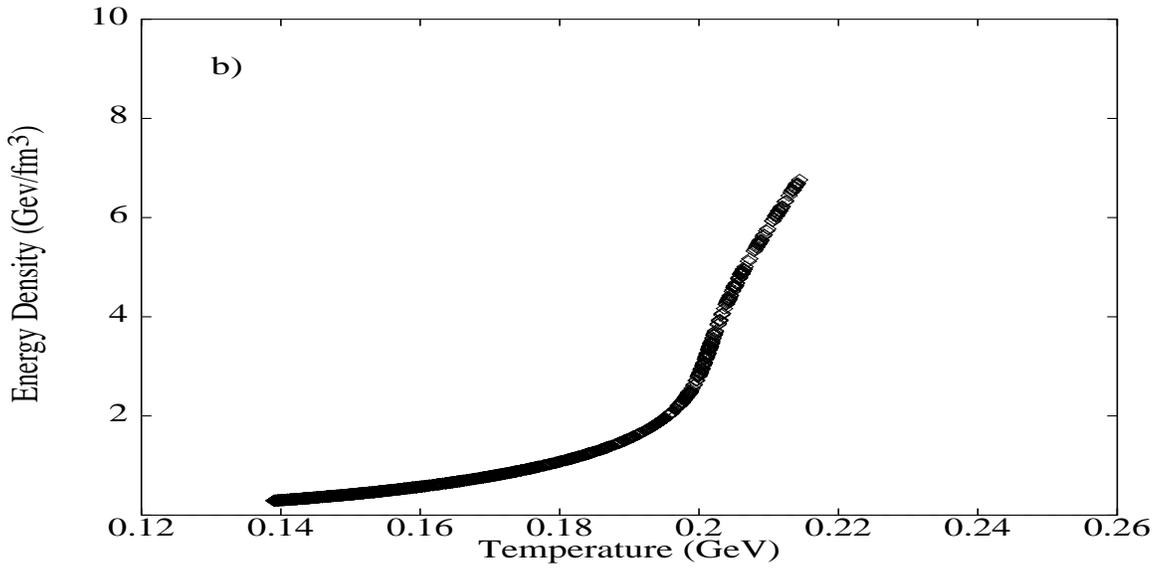
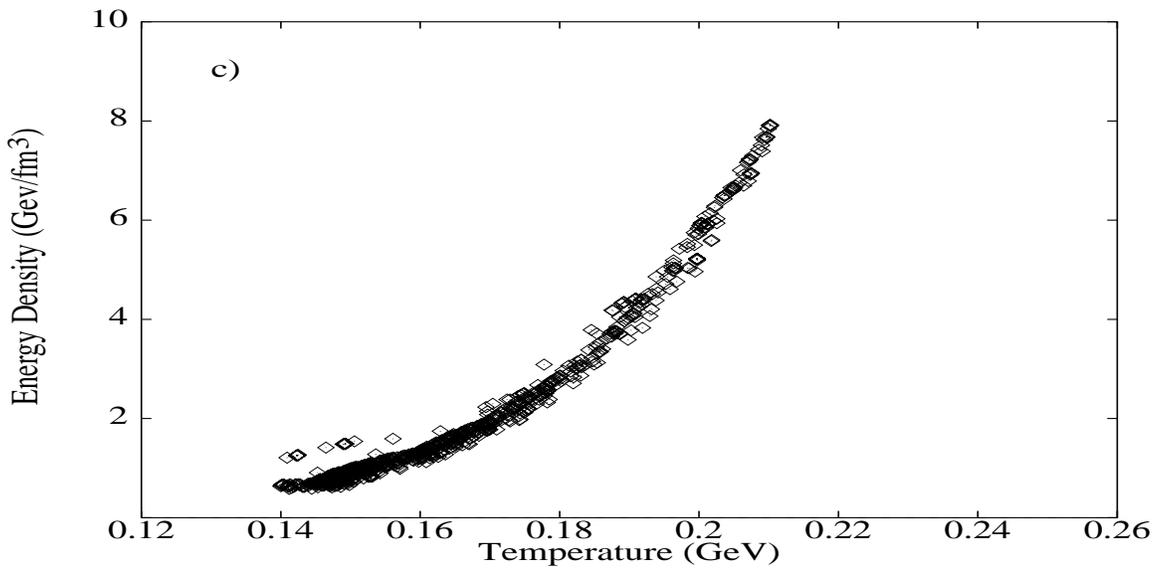
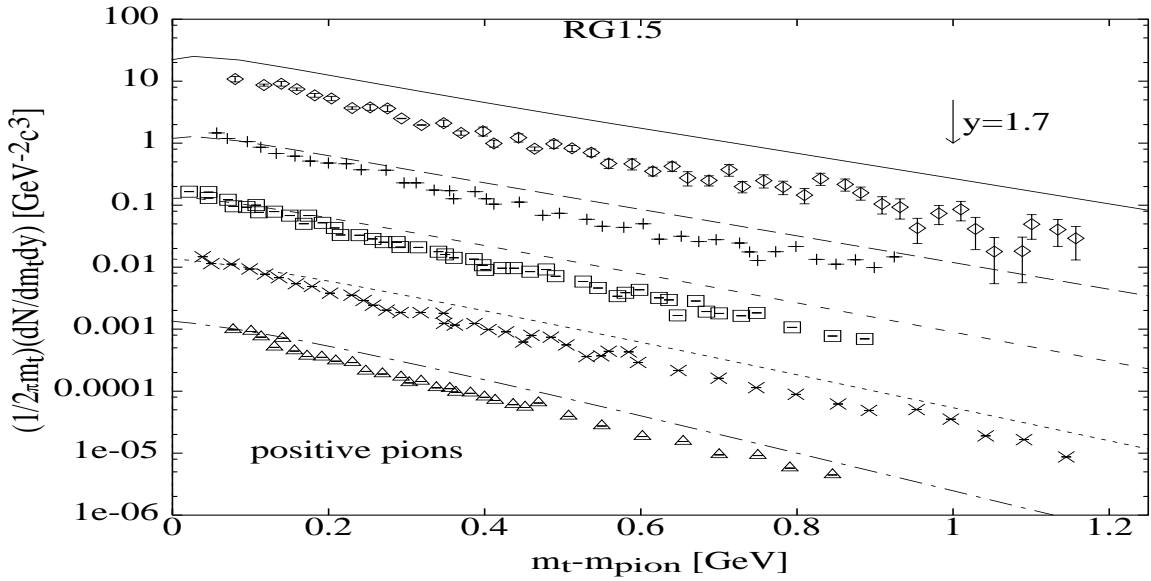
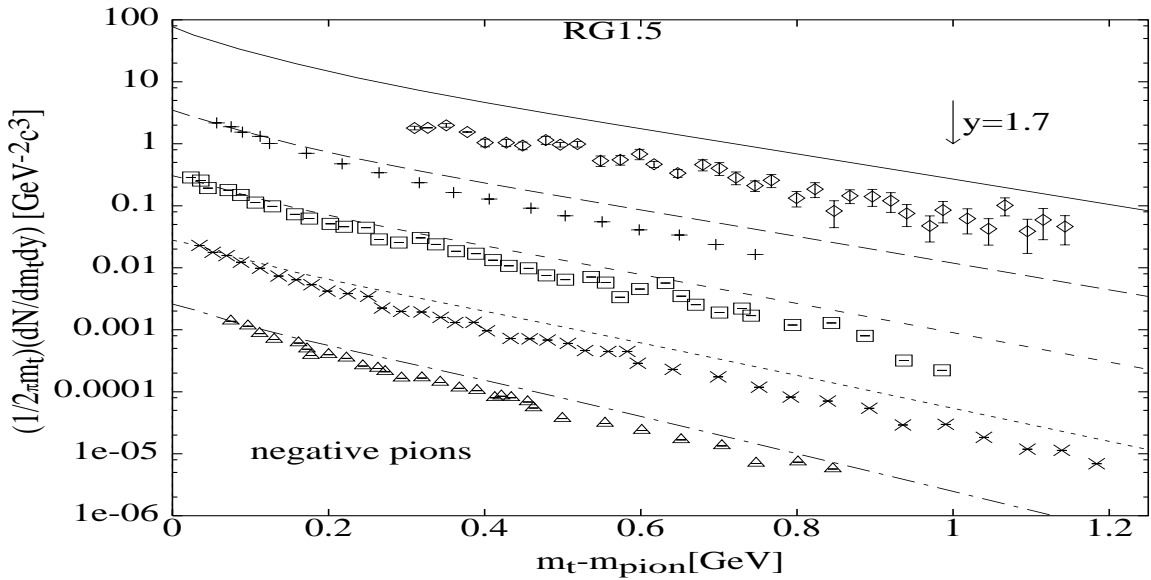
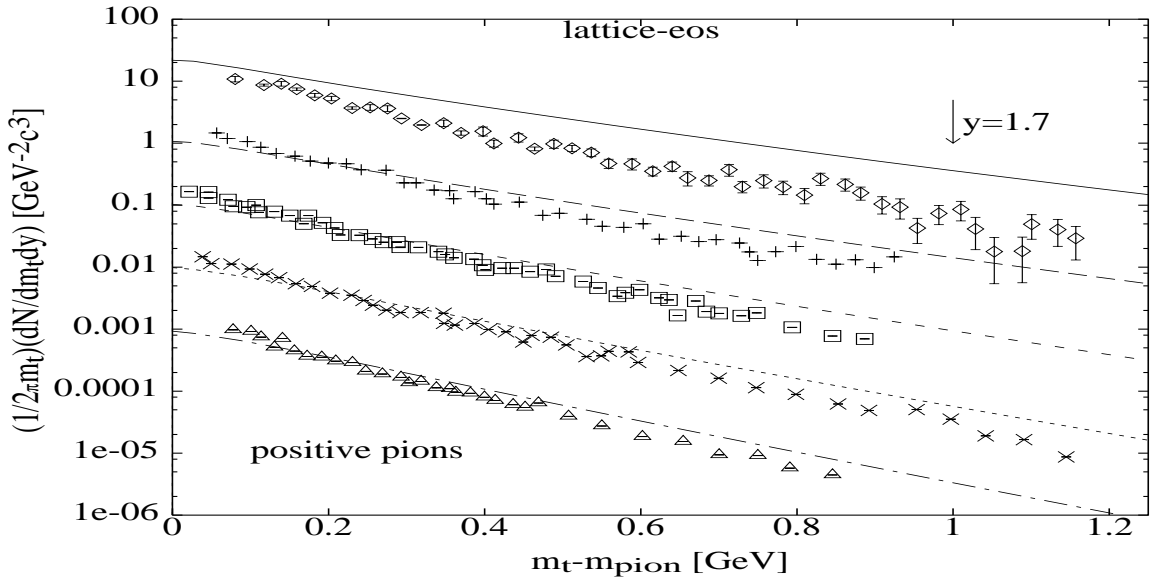
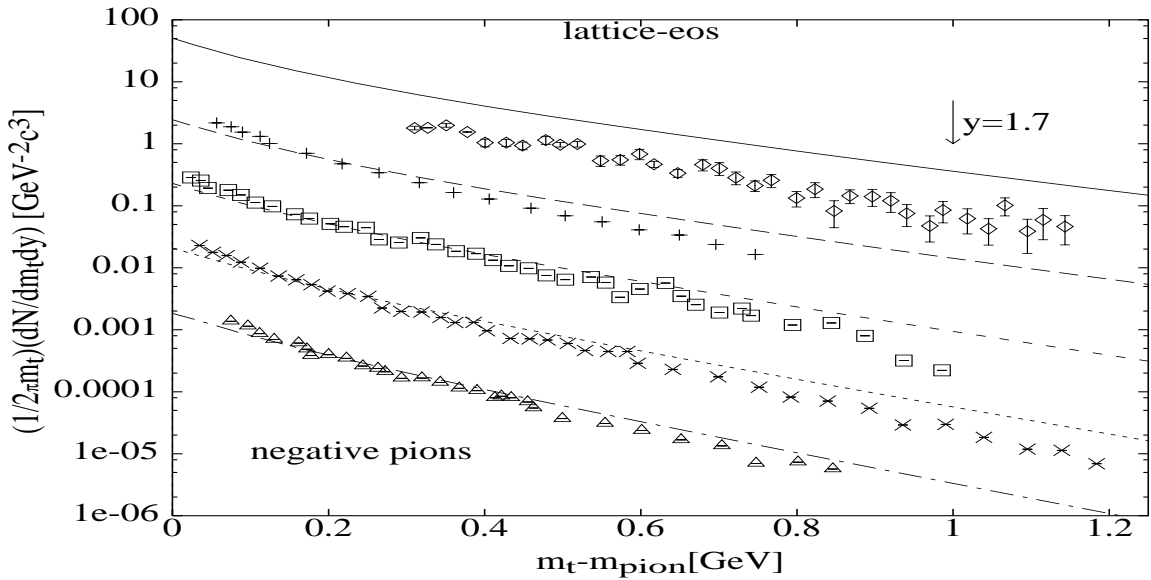
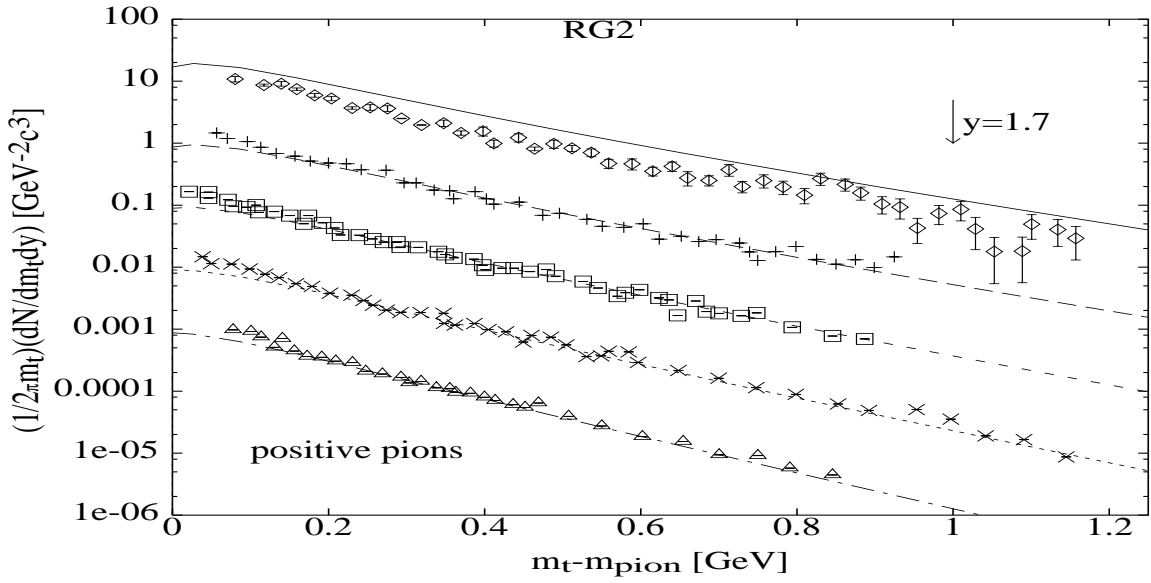
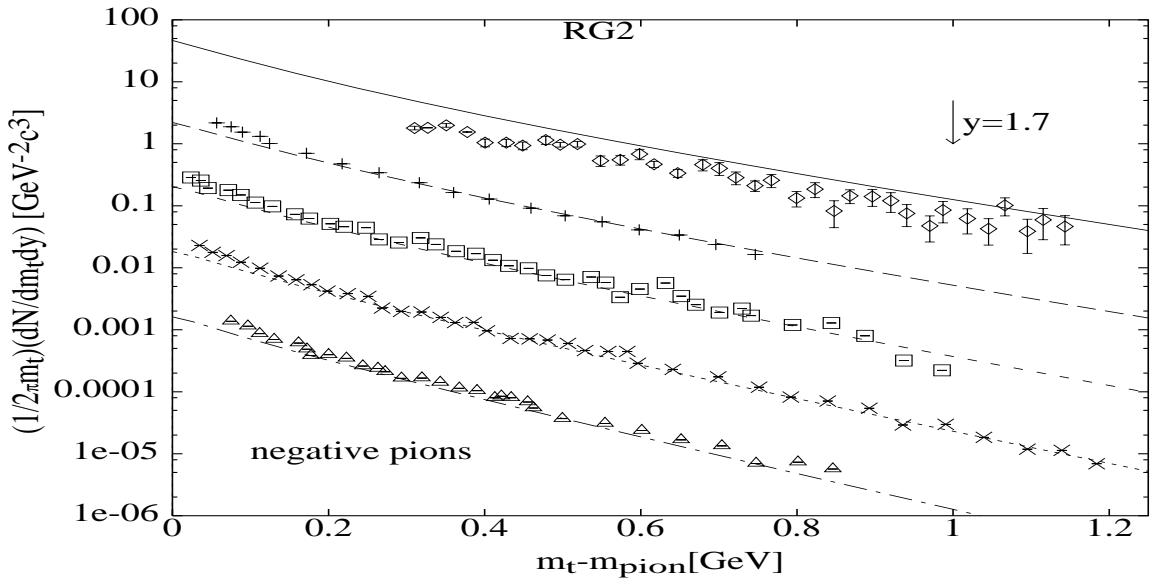
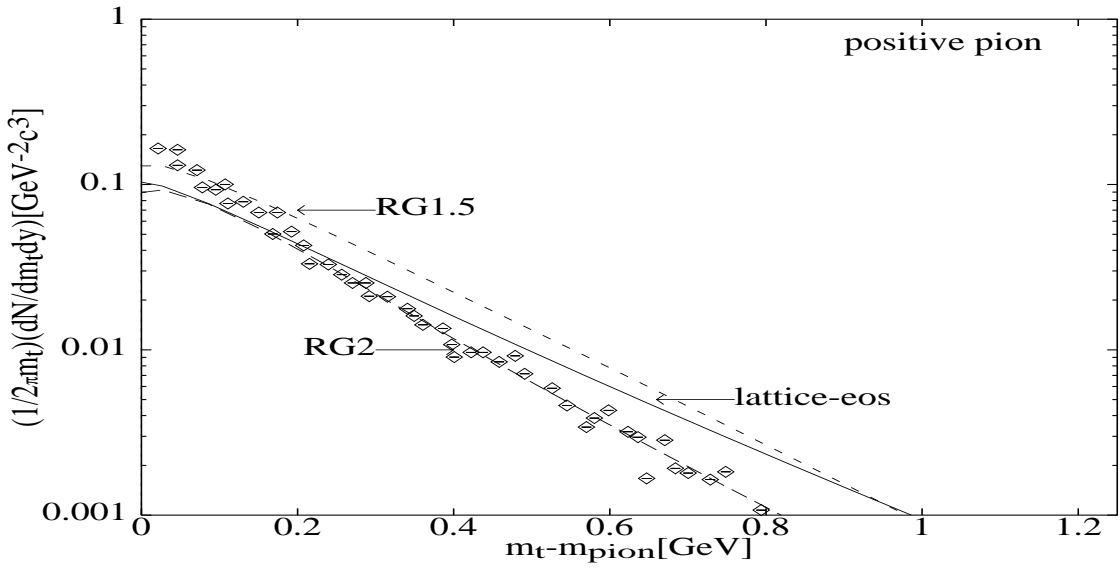
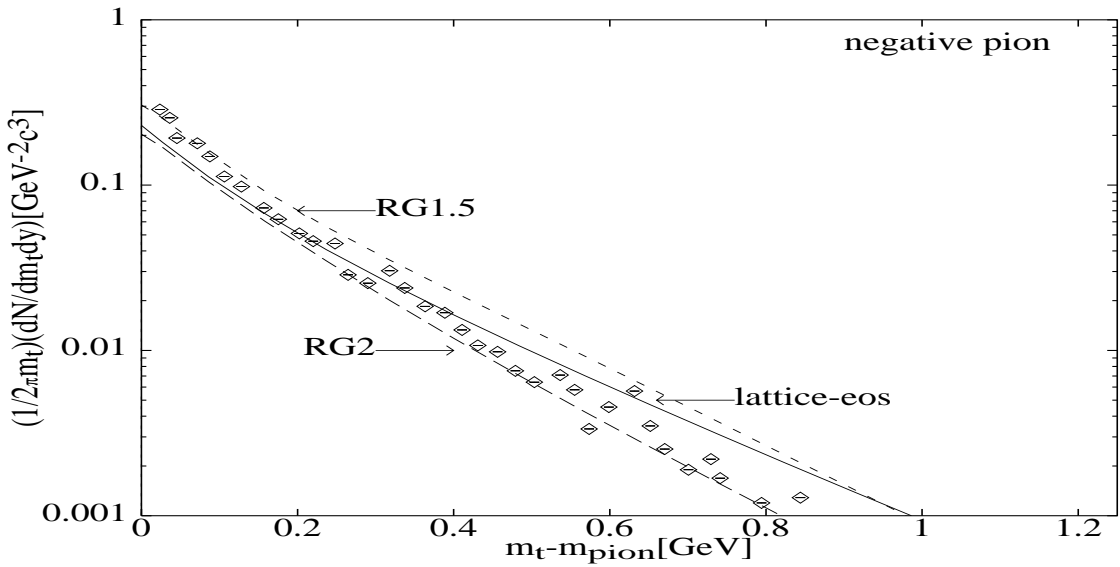
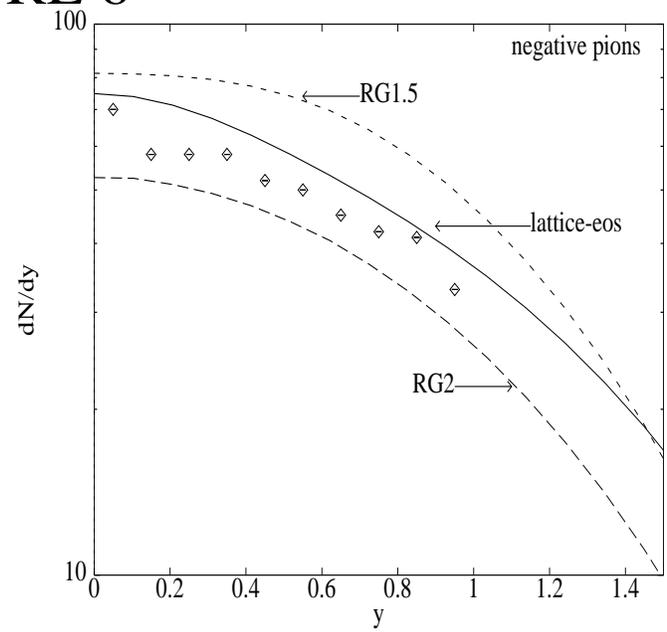
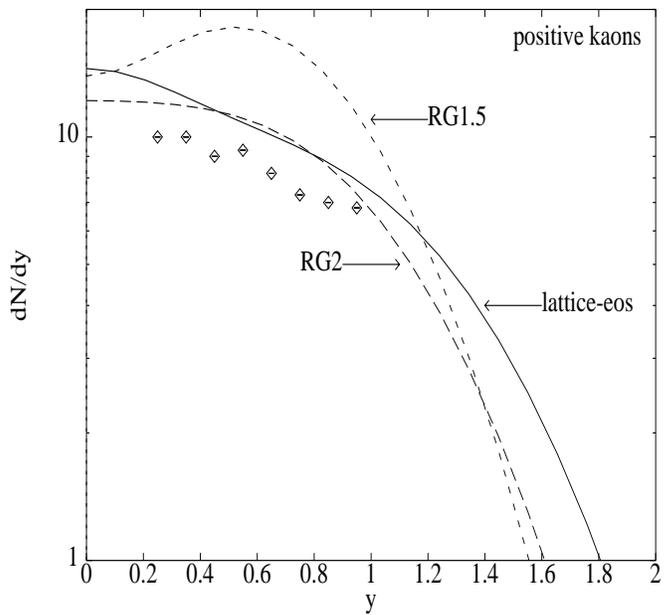
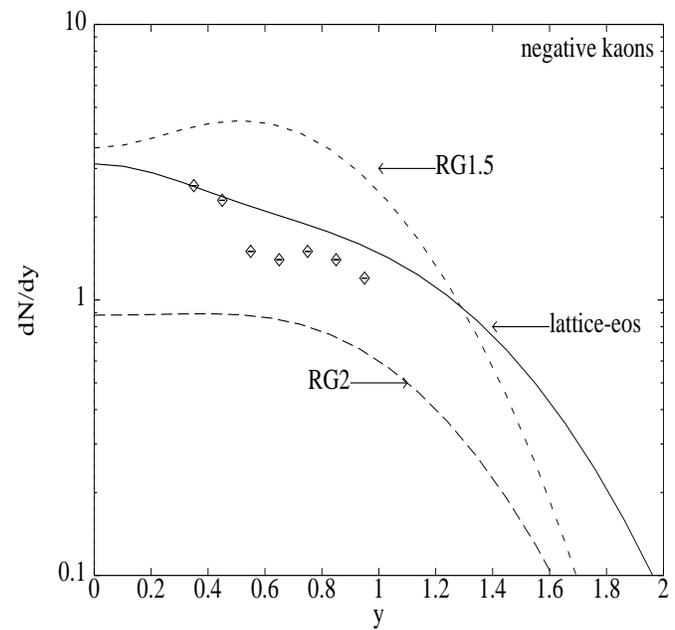
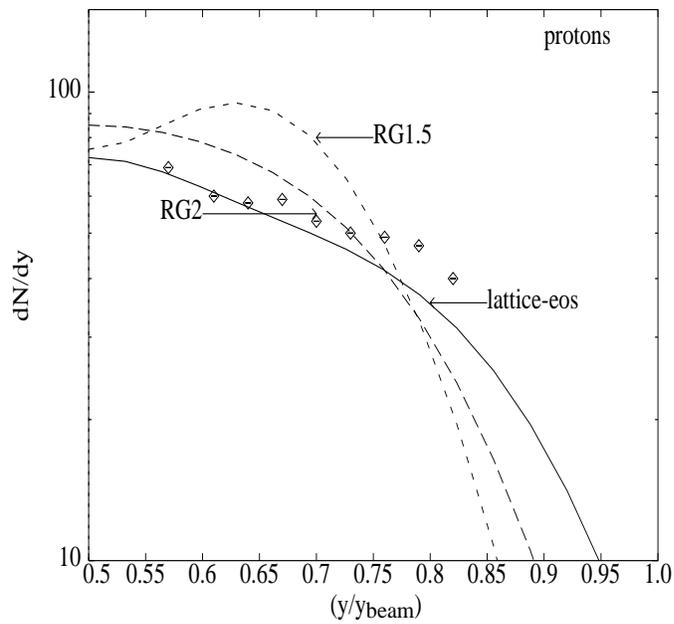
\end{document}